# Synthesis and characterization of bulk Nd$_{1-x}$Sr$_x$NiO$_2$ and Nd$_{1-x}$Sr$_x$NiO$_3$

Bi-Xia Wang[1], Hong Zheng[1], E. Krivyakina[1,2], O. Chmaissem[1,2], Pietro Papa Lopes[1], J.W. Lynn[3], Leighanne C. Gallington[4], Y. Ren[4], S. Rosenkranz[1], J.F. Mitchell[1], and D. Phelan[1]

[1] *Materials Science Division, Argonne National Laboratory, Lemont, IL 60439, USA*
[2] *Department of Physics, Northern Illinois University, De Kalb, IL 60115, USA*
[3] *NIST Center for Neutron Research, National Institute of Standards and Technology, Gaithersburg, MD 20899, USA*
[4] *Advanced Photon Source, Argonne National Laboratory, Lemont, IL 60439, USA*

## Abstract

The recent reports of superconductivity in Nd$_{1-x}$Sr$_x$NiO$_2$/SrTiO$_3$ heterostructures have reinvigorated interest in potential superconductivity of low-valence nickelates. Synthesis of Ni$^{1+}$-containing compounds is notoriously difficult. In the current work, a combined sol-gel combustion and high-pressure annealing technique was employed to prepare polycrystalline perovskite Nd$_{1-x}$Sr$_x$NiO$_3$ ($x$ = 0, 0.1 and 0.2). Metal nitrates and metal acetates were used as starting materials, and the latter were found to be superior to the former in terms of safety and reactivity. The Nd$_{1-x}$Sr$_x$NiO$_3$ compounds were subsequently reduced to Nd$_{1-x}$Sr$_x$NiO$_2$ using calcium hydride in a sealed, evacuated quartz tube. To understand the synthesis pathway, the evolution from NdNiO$_3$ to NdNiO$_2$ was monitored using *in-situ* synchrotron *X*-ray diffraction during the reduction process. Electrical transport properties were consistent with an insulator-metal transition occurring between $x$ = 0 and 0.1 for Nd$_{1-x}$Sr$_x$NiO$_3$. Superconductivity was not observed in our bulk samples of Nd$_{1-x}$Sr$_x$NiO$_2$. Neutron diffraction experiments at 3 K and 300 K were performed on Nd$_{0.9}$Sr$_{0.1}$NiO$_2$, in which no magnetic Bragg reflections were observed, and the results of structural Rietveld refinement are provided.

## 1. Introduction

Since the discovery of superconductivity in cuprates, researchers have targeted low-valence nickelates with square planar coordination because they mimic the cuprates in terms of the electronic count and orbital polarization [1–3]. The infinite layer compounds, $R$NiO$_2$ ($R$ = La, Nd), represent perhaps the simplest such structural prototype and were initially investigated as oxygen deficient phases of $R$NiO$_3$ [4–7]. Despite efforts that span nearly three decades of research on these materials, only recently did the first report of superconductivity appear, in which it was observed up to ~ 15 K in Nd$_{1-x}$Sr$_x$NiO$_2$ thin films grown on SrTiO$_3$ substrates [8]. This intriguing result has rekindled interest in the infinite-layer $R$NiO$_2$ compounds, and a number of theoretical studies have been carried out to investigate the electrical structure of these compounds [9–13]. Calculations indicate that although $R$NiO$_2$ compounds possesses larger charge transfer energies than cuprates [14,15], they do possess similar long-range to nearest neighbor hopping ratios and $e_g$ orbital splitting to the cuprates [12]. Unlike cuprates, $R$ 5$d$ states cross the Fermi level [12,15]. Electronic calculations suggest that the parent compounds (i.e. the undoped, Ni$^{1+}$ end members) should be magnetically ordered; yet, earlier reports by Hayward *et al.* for LaNiO$_2$ [16] and NdNiO$_2$ [17] evidenced no magnetic order via neutron scattering.

Despite the large number of recent theoretical investigations into $R$NiO$_2$ compounds, experimental reports are far more limited. Li *et al.* recently provided evidence for a superconducting dome in Nd$_{1-x}$Sr$_x$NiO$_2$ persisting from 0.125 < $x$ <0.25 and surrounded by weakly insulating states at higher and lower $x$ [18]. Similar results were recently reported by Zeng *et al*. [19]. Lee *et al*. discussed the nuanced complexities of the epitaxial growth [18]. Hepting *et al*. measured *X*-ray absorption spectroscopy and resonant inelastic *X*-ray scattering, interpreting that the rare earth 5*d* states hybridize with nickel 3$d_{x2-y2}$ states [15]. There have been no reports of superconductivity in the bulk, but rather, it has been suggested that the superconductivity may be related to an interface effect [20]. Beyond these reports, little has been experimentally evidenced, and the relative dearth of experimental results is most certainly tied to the challenges presented by synthesis.

Concerning the *bulk R*NiO$_2$ phase, the only reported approach to synthesize it is to prepare polycrystalline samples of Nd$_{1-x}$Sr$_x$NiO$_3$ and then reduce them to Nd$_{1-x}$Sr$_x$NiO$_2$ via a reducing agent [4,5,16]. However, the formation of Nd$_{1-x}$Sr$_x$NiO$_3$ from NiO involves the oxidation of nickel from Ni$^{2+}$ to Ni$^{3+x}$, necessitating the use of high oxygen pressures during synthesis [10]. We note that generally higher pressures are required for smaller $R^{3+}$ species, so that $R$ = Nd growth is more difficult than $R$ = La growth. Although it is unclear what pressures would be required under standard solid-state reaction to produce stoichiometric Nd$_{1-x}$Sr$_x$NiO$_3$ as a function of $x$, it is certain that the synthesis of Sr$^{2+}$ doped NdNiO$_3$ would be more challenging because the introduction of Sr$^{2+}$ to NdNiO$_3$ forces the oxidation state of nickel to exceed 3+. The first report of NdNiO$_3$ was via solid state reaction, which required high oxygen pressure of 60 kbar and high temperature of 950 °C [10]. However, an alternative route for producing polycrystalline samples of NdNiO$_3$ was later developed by Lacorre *et al.* through sol-gel precursors by using nitrates as starting materials [5,10]. The advantage of the sol-gel route is that precursor powders are ultra-fine so that their reactivity is much higher than that of binary oxides and carbonates at reduced temperatures due to the short diffusion path of the ions [11,21]; thus, in principle, synthesis can be carried out at lower temperatures and pressures. Nevertheless, for perovskite nickelates, while this approach does reduce the required pressure, it still involves heat treatment at temperatures greater than 1000 °C and under oxygen pressures as high as 200 bar [22].

Given the recent surge of interest in finding superconductivity in the $R$NiO$_2$ phase, we have investigated the synthesis of Nd$_{1-x}$Sr$_x$NiO$_3$ precursors as well as the reduction process of Nd$_{1-x}$Sr$_x$NiO$_3$ to Nd$_{1-x}$Sr$_x$NiO$_2$. Here, we provide details of the procedures that we have used to synthesize the materials, comparing different sol-gel combustion methods for Nd$_{1-x}$Sr$_x$NiO$_3$, performing *in-situ* X-ray diffraction during the reduction process to obtain NdNiO$_2$, and discussing the challenges in obtaining high purity, high quality materials. We characterize the electrical transport properties of the obtained materials, and provide structural Rietveld refinements of neutron diffraction data carried out on Nd$_{1-x}$Sr$_x$NiO$_2$ at 3 K and 300 K.

## 2. Experimental

## 2.1 Synthesis of $Nd_{1-x}Sr_xNiO_3$

**Nitrate method**: $Nd_{1-x}Sr_xNiO_3$ ($x$ = 0, 0.1 and 0.2) was synthesized via the citrate-nitrate auto-combustion synthesis as described by Deganello *et al*. [23] Metal nitrates $Nd(NO_3)_3 \cdot 6H_2O$ (Strem Chemicals, 99.999%), $Ni(NO_3)_2 \cdot 6H_2O$ (Sigma-Aldrich, 99.999%) and $Sr(NO_3)_3$ (Alfa Aesar, 99.97%) were used as starting materials. The metal nitrates were dissolved in a minimum amount of distilled water to obtain a clear solution. The exact concentrations of metal nitrates were determined by inductively coupled plasma mass spectrometry (ICP-MS) to account for variable levels of hydration in the starting materials. Stoichiometric amounts of nitrates were mixed in a beaker to get the desired $Nd_{1-x}Sr_xNiO_3$ ($x$ = 0 to 0.2) precursor solutions. An aqueous solution of citric acid was then mixed with the precursor solutions at a molar ratio, citric acid to metal ions, of 1.2:1 at room temperature. The pH of the precursor solution was adjusted to ~ 7 by adding 30 wt. % aqueous $NH_3$. The solution was slowly evaporated at 70 °C with continuous mechanical stirring, until a hardened gel was obtained. To carry out the gel decomposition under controlled conditions, the hardened gel was transferred to an alumina crucible, which was then put on a hot plate at 250 – 300 °C. The gel first turned to black, and then ignited and underwent a vigorous self-sustaining combustion, yielding an ash product. The combusted powders were then fired at 800 °C for 12 h in an oxygen atmosphere at ambient pressure with a flow rate of 0.2 L/min. The resulting black powder was then ground and pressed into pellets with dimensions of 13 mm diameter and about 0.2 mm thickness. The pellets were subsequently fired at 1000 °C under 150-160 bar of oxygen pressure for 12 hours in a high-pressure annealing furnace (Model AHSO, SciDre GmbH Dresden). For $x$ = 0.10 and 0.20, multiple 12 hours firings with intermediate regrinding were carried out.

**Citrate method**: $Nd_{1-x}Sr_xNiO_3$ (0.0 ≤ $x$ ≤ 0.2) was also synthesized using metal acetates, $Nd(CH_3CO_2)_3 \cdot H_2O$, $Ni(CH_3CO_2)_2 \cdot 4H_2O$ and $Sr(CH_3CO_2)_3$, as starting materials. To prevent the formation of carbonate precipitation over time, the pH of the acetate solutions was adjusted to 1 by addition of $HNO_3$. The rest of the experiment procedure was similar to that described for metal nitrates. Note that the combustion reaction using citrates as starting materials was much less vigorous compared to the one using nitrates.

## 2.2 Reduction of $Nd_{1-x}Sr_xNiO_3$

The reduction of $Nd_{1-x}Sr_xNiO_3$ (0.0 ≤ $x$ ≤ 0.2) was performed using $CaH_2$ powder as a reducing agent. From the previous obtained $Nd_{1-x}Sr_xNiO_3$ (0.0 ≤ $x$ ≤ 0.2), 90 - 100 mg chunks extracted from the thin pellets were wrapped loosely with aluminum foil, which had small holes in it. Then the wrapped samples were loaded together with 0.25g $CaH_2$ powder in an Ar-filled glove box and subsequently sealed under rough vacuum ($10^{-3}$ mbar) in an evacuated quartz tube with 11 mm outer diameter, 9 mm inner diameter, and approximately 200 mm length. The tube was then heated to 285 °C at a rate of 1 °C/min and kept at this temperature for 48 hours. After reduction, the remaining $CaH_2$ powder on the sample surfaces was gently removed.

## 2.3 Characterization

Powder *X*-ray diffraction data were collected at room temperature on all samples using a PANalytical X'Pert Pro powder diffractometer with Cu Kα radiation ($\lambda$ =1.5418 Å) in the 2$\theta$ range of 20-70º. Rietveld refinement was performed on the collected data using GSAS II software [24].

*In-situ* high energy synchrotron *X*-ray diffraction experiments were conducted at beamline 11-ID-C ($\lambda$ = 0.11173 Å, beam size = 0.5 mm × 0.5 mm), Advanced Photon Source (APS) at Argonne National Laboratory. The sample chunk of $NdNiO_3$ (2 mg), and $CaH_2$ powder (50 mg) were loaded in a Pyrex glass tube with 1.25 mm inner diameter (ID), and then the loaded tube was sealed under vacuum of $10^{-3}$ mbar. A specially designed electrical resistance heating furnace was used to heat the sample. The heating cycle was as follows: ramping at 10 ˚C/min to 200 ˚C, and then at 1.6 ˚C/min to 340 ˚C, holding at 340 ˚C for 120 min, then cooling to room temperature at 10 ˚C/min. Synchrotron powder diffraction data were collected at 5 minute intervals.

Resistivity measurements of the bulk $Nd_{1-x}Sr_xNiO_3$ ($0.0 \leq x \leq 0.2$) and $Nd_{1-x}Sr_xNiO_2$ ($0.1 \leq x \leq 0.2$) samples were performed in a Quantum Design physical property measurement system (PPMS) in the temperature range of 2–300 K using the Van der Pauw method with contacts made with silver paste.

Powder neutron diffraction experiments were performed on the BT-1 diffractometer at the NIST Center for Neutron Research using the Ge (311) monochromator with $\lambda$ = 2.079 Å. The mass of the sample was ~1 g, was sealed in a vanadium can under helium atmosphere and placed in a closed-cycle refrigerator with a base temperature of 3 K.

## 3. Results

### 3.1 Synthesis of $Nd_{1-x}Sr_xNiO_3$ ($x$ = 0, 0.1 and 0.2)

The combustion synthesis of citric acid with nitrates produces ultrafine powders, which is confirmed by the size-broadened *X*-ray diffraction pattern shown in Fig.1. These as-combusted powders were then annealed in an $O_2$ flowing atmosphere at ambient pressure at 800 °C for 12 hours, after which it was found that the majority phase of the powder was already $NdNiO_3$ with a small amount of $Nd_2NiO_4$ and some unreacted NiO and $Nd_2O_3$ phases (Fig.1). The black powder was then finely ground and fired again under the same conditions for 84 hours. As shown in Fig.1, some $Nd_2NiO_4$ and NiO remained, but the quantity of impurity phases was significantly reduced. Because of the poor crystallinity, evidenced by the broadened reflections, and small amount of second phases, the sample was then fired at 1000 °C for 12 hours under an oxygen pressure of 150-160 bar. As shown in Fig.2(a), the crystallinity and purity of the resulting powder was improved markedly over that reached under ambient conditions.

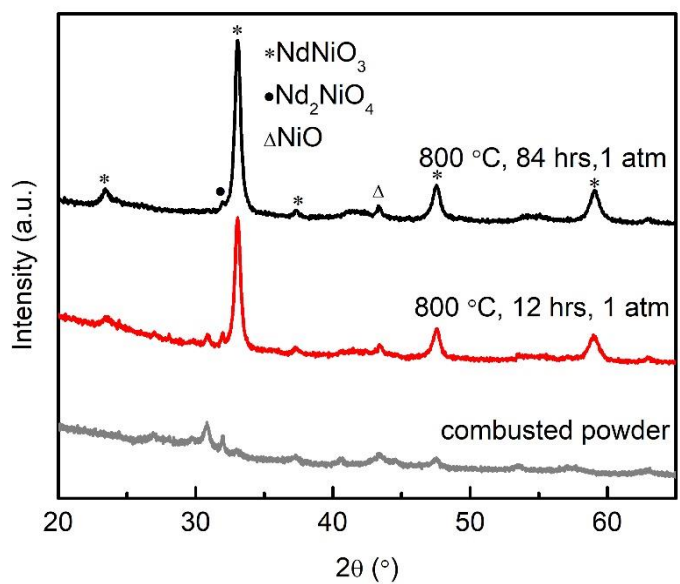

**Fig. 1**: *X*-ray powder diffraction patterns of combusted Nd$_{1-x}$Sr$_x$NiO$_3$ ($x = 0$) powder from nitrate method and after annealing in O$_2$ at ambient pressure for 12 hours and 84 hours respectively.

The incorporation of Sr$^{2+}$ into NdNiO$_3$ increases the formal nickel valence and consequently requires additional high-pressure steps. The main secondary phase involved in the synthesis for both $x = 0.1$ and 0.2 samples is (Nd,Sr)$_2$NiO$_4$ (214), which is more stable at low oxygen pressure. For $x = 0.1$, a firing at 1000 °C for 12 hours under O$_2$ pressure of 150 bar was needed to remove the 214 impurity, whereas for $x = 0.2$, multiple firings at 1000 °C under 150 bar were required to suppress the secondary phase. As shown in Fig. 2(b) and (c), both $x = 0.1$ and 0.2 samples were found to contain small quantities of NiO impurities.

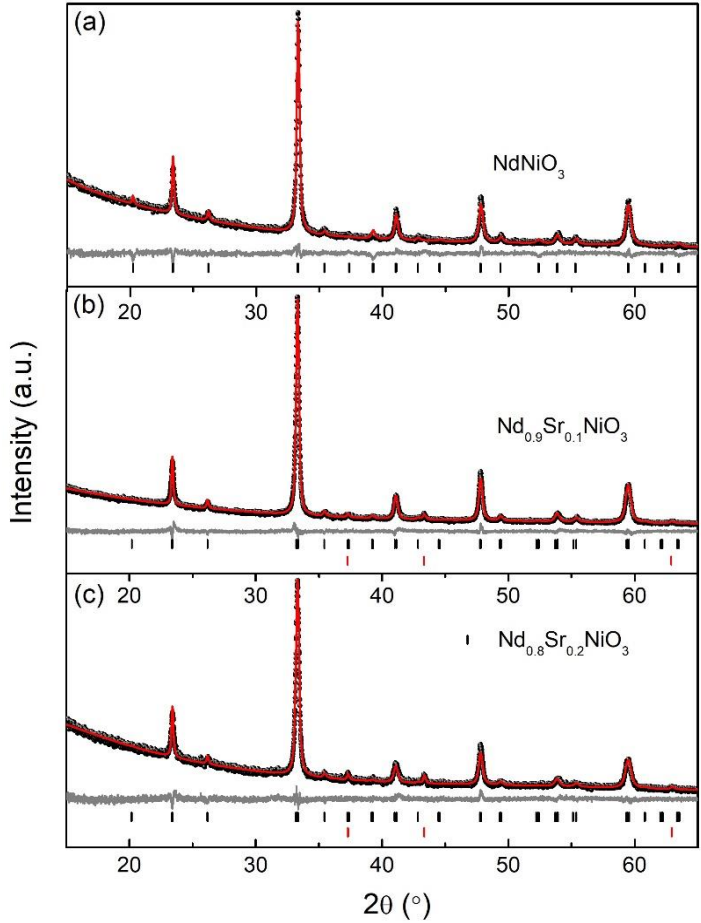

**Fig. 2**: Room temperature *X*-ray powder diffraction patterns of $Nd_{1-x}Sr_xNiO_3$ ($0.0 \leq x \leq 0.2$) and corresponding Rietveld fitting results using *Pbnm* space group. The black dots are the observed data, the red line is the calculated fit, and gray line shows the difference between the two. Positions of allowed reflections are indicated by vertical black ((Nd,Sr)$NiO_3$) and red (NiO) lines.

All three compositions were indexed in the orthorhombic in *Pbnm* space group (Fig. 2), and Rietveld refinement was carried out. The refined parameters and lattice constants are listed in Table 1. The variation of the lattice parameters and volume with the Sr content is plotted in Fig. 3. The good agreement factors are consistent with a random substitution of $Sr^{2+}$ for the rare-earth cations. As shown in Fig. 3 (a), the lattice parameter *c* increases with Sr content while *b* decreases, and *a* is practically constant over this doping range. The trends in *b* and *c* are consistent with those previously reported by Alonso et al. [25], and the values of *a* are similar although Alonso et al. reported a monotonic increase in *a*.

The observed unit cell volume (Fig. 3(b)) varies non-monotonically, with a minimum appearing at *x* = 0.1, though the change in volume is very small (0.1%). Alonso et al. also observed a non-monotonic behavior, with small changes, though the minimum was observed at x = 0.5 [25]. Given the combination of statistical and systematic uncertainty, it is our contention that the volume is nearly constant with substitution in this doping range and that the non-monotonic behavior should be considered as a possible, but not definitive trend. Based on the ionic size of the A-site dopant alone ($Nd^{3+} \approx 1.27$ Å, $Sr^{2+} \approx 1.44$ Å) [26], one should expect an increase in the unit cell volume in a Vegard's law-like fashion. However, additional complications could arise from the increasing propensity for oxygen vacancies to form, which would also be expected to increase the cell volume, and the change in the nickel valence, which would be anticipated to decrease the cell volume. Furthermore, reciprocal trends between lattice constants (e.g., $b$ decreasing, $c$ increasing) can also reflect the perovskite's propensity to compensate for ionic size effects through rotations of oxygen octahedra. Given the imprecision of *X*-ray diffraction in determining oxygen positions, we do not further elaborate on this last point.

**Table 1 Lattice parameters of $Nd_{1-x}Sr_xNiO_3$ (x = 0.0, 0.1 and 0.2) compounds made from nitrates determined by Rietveld refinement of *X*-ray data.**

| $x$ | $a$ (Å) | $b$ (Å) | $c$ (Å) | $V$ (Å$^3$) | GOF | Rw |
|---|---|---|---|---|---|---|
| 0.0 | 5.3958(8) | 5.3758(8) | 7.6034(8) | 220.55(5) | 1.55 | 5.70% |
| 0.1 | 5.394(2) | 5.368(2) | 7.609(2) | 220.32(2) | 1.41 | 4.87% |
| 0.2 | 5.3963(5) | 5.3647(5) | 7.6133(8) | 220.40(3) | 1.21 | 5.14% |

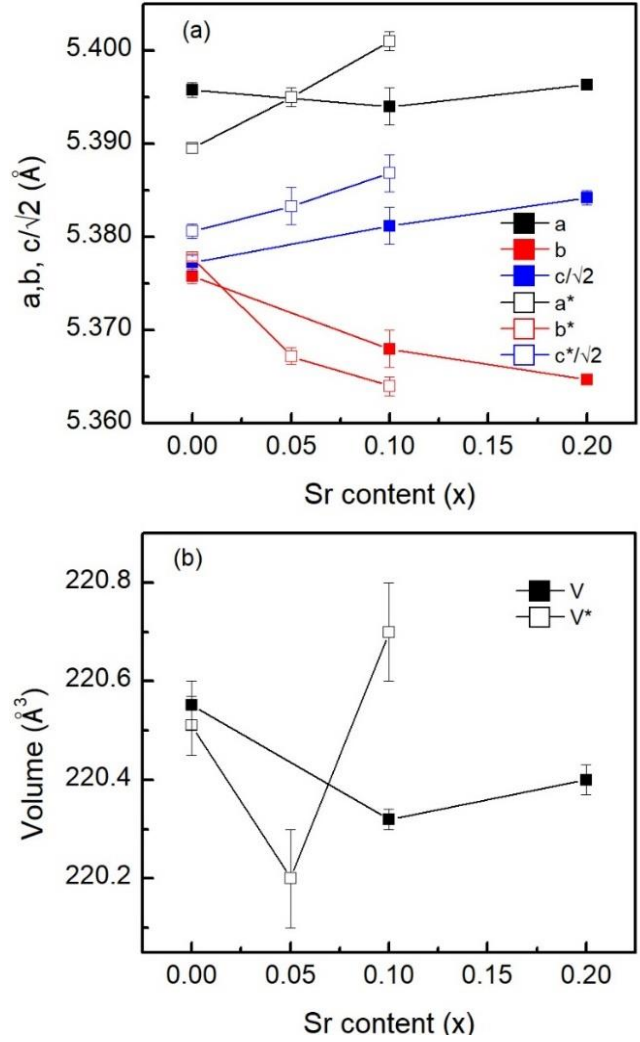

**Fig. 3**: Variation of the (a) unit-cell parameters and (b) cell volume of $Nd_{1-x}Sr_xNiO_3$ as a function of the Sr content ($x$). Solid squares: this work, open squares from Ref. [25]. Uncertainties represent on standard deviation.

### 3.2 Acetates vs. nitrates

Due to the vigor of the reaction using metal nitrates, only approximately 1 g of the hardened gel was burned at a time. In some publications [27,28] for the preparation of battery materials via sol-gel method, metal acetates were also employed as oxidants. Therefore, $Nd_{1-x}Sr_xNiO_3$ ($x = 0, 0.1, 0.2$) was also synthesized using metal acetates $Nd(CH_3CO_2)_3 \cdot H_2O$, $Ni(CH_3CO_2)_2 \cdot 4H_2O$ and $Sr(CH_3CO_2)_3$ in a manner similar to that described using nitrates. Combustion of larger quantities (~ 5 gram) could be conducted each time thanks to the less energetic reaction. Here we take the synthesis of $Nd_{0.9}Sr_{0.1}NiO_3$ using metal acetates as an example. As shown in Fig. 4(a), the combusted powder from metal acetates shows no Bragg reflections at all, while a few peaks from metal oxide were present in the product from the metal nitrates. After annealing at 800 °C for 12 hours at ambient conditions, the majority of the sample made from metal acetates was already Nd$_{0.9}$Sr$_{0.1}$NiO$_3$. In the case of the sample made from nitrates, there was still about 40% of (Nd,Sr)$_2$NiO$_4$ phase estimated from the *X*-ray (Fig. 4(b)). This suggests that the reactivity of the powder obtained from the metal acetates combustion reaction was higher. In the end, Nd$_{0.9}$Sr$_{0.1}$NiO$_3$ was achieved after one single firing at 1000 °C for 12 h in both cases under an oxygen pressure of 150 bar (Fig. 4(c)). However, as shown in the Fig.4(d), the full-width-half-maximum (FWHM) for the product from metal acetates (0.261°) is slightly smaller than that of the sample from metal acetates (0.291°), which indicates that the *X*-ray quality of the final products from metal acetates is also slightly better than that of from metal nitrates.

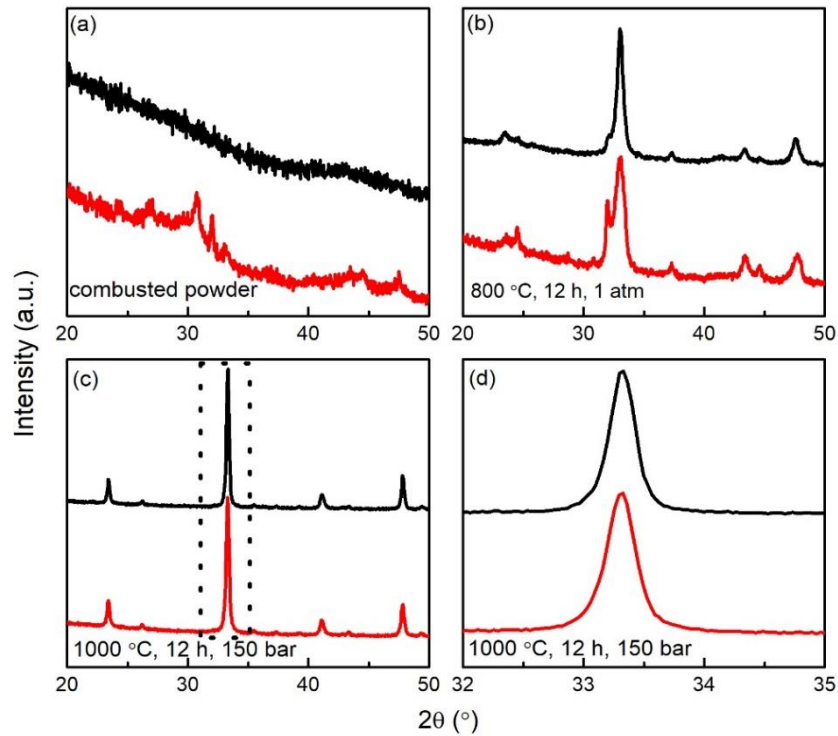

**Fig. 4**: *X*-ray powder diffraction patterns for samples made from acetates (red) and nitrates (black). (a) Combusted powders, (b) powders annealed at 800 °C for 12 hours under ambient pressure, and (c) the data annealed at 1000 °C for 12 hours under oxygen pressure of 150 bar. (d) Enlarged view of the region marked by rectangle in (c).

### 3.2 Synthesis of Nd$_{1-x}$Sr$_x$NiO$_2$ by reducing Nd$_{1-x}$Sr$_x$NiO$_3$

As discussed above, the RNiO$_2$ phase can be obtained by reduction of the RNiO$_3$ phase. Previous studies have shown that the reduction of NdNiO$_3$ in flowing, diluted H$_2$ gas leads to a reported Nd$_3$Ni$_3$O$_7$ phase [5]. The infinite layer Ni$^{1+}$ containing the NdNiO$_{2+x}$ phase is most easily accessible with the hydride reagents [17]. In the present experiment, the reduction of Nd$_{1-x}$Sr$_x$NiO$_3$ was performed using CaH$_2$ as a reducing agent. As described in the experiment section, 90 - 100 mg chunks extracted from the polycrystalline Nd$_{1-x}$Sr$_x$NiO$_3$ pellets were either directly embedded in CaH$_2$ powders or loosely wrapped with aluminum foil (with holes) before mixing with CaH$_2$. It was found that the results from both methods are identical. In the early work done by Hayward *et al.* [16,17,29–31], in which NaH or CaH$_2$ was used as reducing agent, the samples were thoroughly ground with the hydride and the reduction has been assumed to be a solid state reaction. Later, Kobayashi *et al*. [32] successfully conducted the reduction of $SrFeO_3$ to $SrFeO_2$ with physically separated $CaH_2$, and it was concluded that the reaction was proceeded via the release of $H_2$ by $CaH_2$. Recently, Page *et al*. [33] reduced $CaMn_{0.5}Ir_{0.5}O_3$ using NaH and a gas mediated reaction process was proposed. Here, our results, which are identical for the physically separated and mixed experiments, once again confirms that the reduction using $CaH_2$ works via $H_2$ gas instead of solid state reaction.

Previous work on thin-film $LaNiO_3$ has shown that reduction induces a series of transformation steps: first to brownmillerite $LaNiO_{2.5}$, then to *c*-axis $LaNiO_2$, followed by a reorientation transition to *a*-axis $LaNiO_2$, before subsequent decomposition [5,7,34]. Additionally, a previous study on $NdNiO_3$ reduction indicated that a fluorite defect phase can be introduced on top of the infinite-layer $NdNiO_2$ (001) films under certain annealing conditions [35]. These results hinted that it is difficult to stabilize a uniform, single-crystalline infinite-layer $NdNiO_2$ phase with its formal $Ni^{1+}$ valence. Indeed, for $NdNiO_3$, where $x = 0.0$ in $Nd_{1-x}Sr_xNiO_3$, our attempts to reduce to $NdNiO_2$ using $CaH_2$ were not successful (except for the *in-situ* reduction experiment described below). We tried to optimize the reduction conditions by adjusting reaction time and temperature, but no full transformation to the infinite-layer structure was realized. Given the fact that Hayward *et al.* succeeded in making $NdNiO_2$ by grinding the $NdNiO_3$ with sodium hydride [16], we also tried to grind $NdNiO_3$ with $CaH_2$ for the reduction experiment, but this, too, was unsuccessful.

Though the reduction of $NdNiO_3$ to $NdNiO_2$ was not successful in our ex-situ experiments, we were able to achieve successful reduction via an *in-situ* high energy synchrotron *X*-ray diffraction experiment. Fig. 5 shows the evolution of phases and the reducing reaction pathway with temperature. As shown in Fig. 5(a), the $NdNiO_3$ phase diminished rapidly at 200˚C along with the appearance of the $Nd_3Ni_3O_7$ phase (purple curve). As the temperature was increased, $NdNiO_2$ began to appear as a secondary phase (Fig. 5(b)), and eventually $Nd_3Ni_3O_7$ did give rise to $NdNiO_2$ at 260˚C (black curve), which was stable until the conclusion of the reduction at 340˚C. This suggests that it is possible to prepare $NdNiO_2$ using $CaH_2$, but it requires subtleties we don't yet understand. One speculation is that the tube size for the *in-situ* experiment is relatively small, which generates a higher effective $H_2$ concentration, but further work is needed to confirm this.

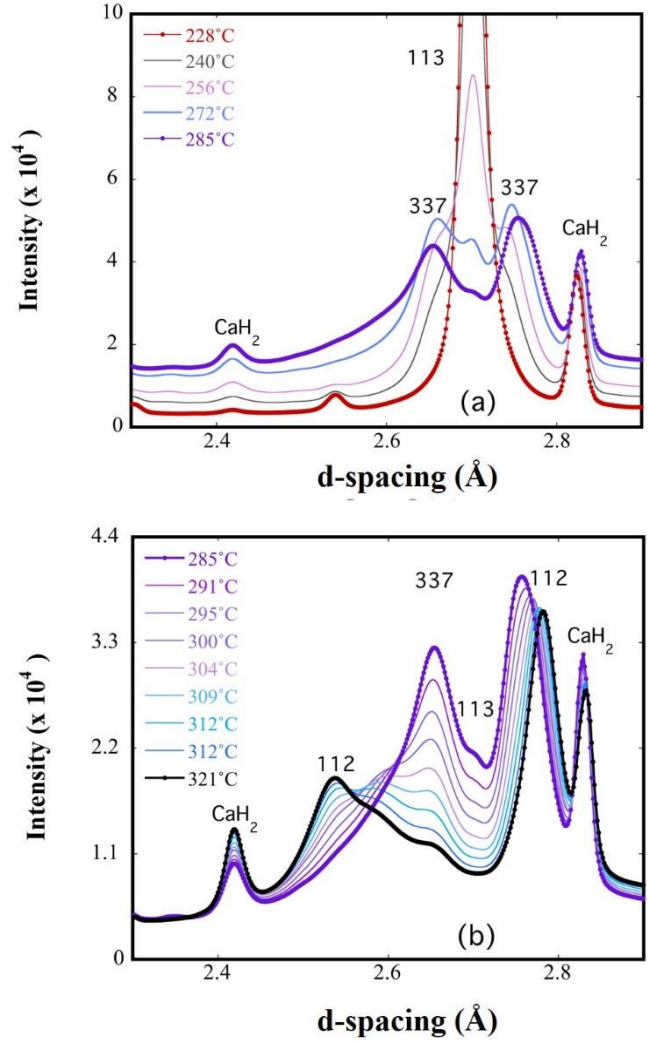

**Fig. 5**: High energy *in-situ* X-ray data showing the evolution from (a) NdNiO$_3$ to Ni$_3$Ni$_3$O$_7$ and (b) Ni$_3$Ni$_3$O$_7$ to NdNiO$_2$. The 113, 112 and 337 denote NdNiO$_3$, NdNiO$_2$ and Nd$_3$Ni$_3$O$_7$ phases, respectively.

Although reduction of the parent phase to NdNiO$_2$ is difficult, we found that hole-doping (increasing *x*) facilitates the reduction process, likely because the formal valence state of nickel in the final product is increased. We were able to successfully reduce $x = 0.1$ and 0.2 samples to Nd$_{1-x}$Sr$_x$NiO$_2$ using the protocol described above, which failed for $x = 0$. Fig. 6 (a) and (b) display the room temperature XRD measurements and the Pawley refinements of Nd$_{1-x}$Sr$_x$NiO$_2$ ($x = 0.1, 0.2$) bulk samples. The main phase of both the samples is the infinite-layer phase with a small amount of a nickel impurity. Nickel metal is a common product for the reduction of perovskite nickelates [16,17] and likely arises from the reduction of NiO impurities that were present in the perovskite phase prior to reduction. This realization highlights the fact that achieving high purity Nd$_{1-x}$Sr$_x$NiO$_2$ requires a high purity perovskite precursor. In fact, we have a dichotomy: it becomes more difficult to grow the perovskite precursor as *x* is increased because of the high nickel valence required (3+*x*), but once the precursor is produced, it is more difficult to reduce the precursor to the infinite later phase as *x* is decreased because of the low valence required $(1+x)$. Considering the challenges involved in both steps, we have found that the easiest infinite layer composition to synthesize has $x \sim 0.1$.

In early work done by Li *et al.*, it was found that the peaks of infinite layer nickelates are broad, which was attributed to randomly oriented domains with different sizes in the sample [20]. In our case, the infinite layer phases are also imperfectly crystallized, and the statistics of our X-ray diffraction data are relatively low, therefore, we only applied the Pawley method for the data refinement. A detailed structural analysis on $Nd_{0.9}Sr_{0.1}NiO_2$ by Rietveld refinement is provided by neutron diffraction below. The calculated lattice parameters for $Nd_{0.9}Sr_{0.1}NiO_2$ and $Nd_{0.8}Sr_{0.2}NiO_2$ by Pawley fitting are $a = 3.9212(7)$ Å, $c = 3.2681(6)$ Å and $a = 3.9175(5)$ Å, $c = 3.2793(4)$ Å, respectively. Specifically, we notice that the in-plane lattice constant $c$ of our $Nd_{0.8}Sr_{0.2}NiO_2$ sample is slightly smaller than that of reported $Nd_{0.8}Sr_{0.2}NiO_2$ thin films (range from 3.34 Å to 3.38 Å) [8] and bulk sample made by Li et al. (3.34 Å) [20]. Ultimately, however, the results from X-ray (3.267 Å) and neutron diffraction (3.268 Å) for $Nd_{0.9}Sr_{0.1}NiO_2$ are internally consistent.

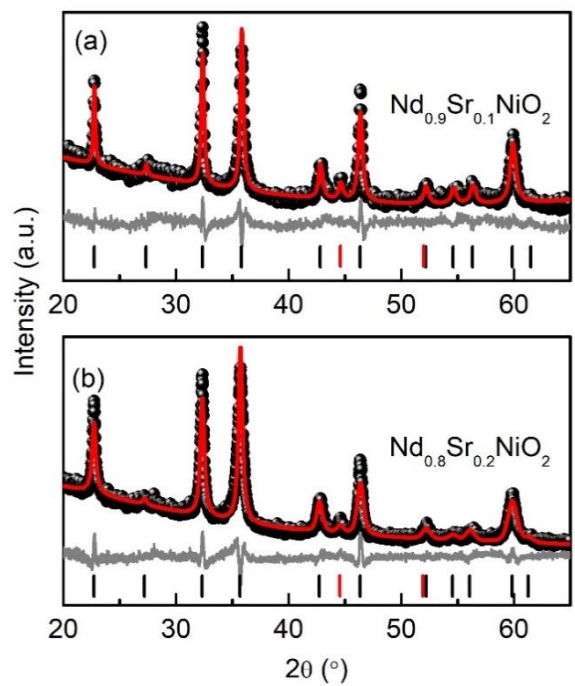

**Fig. 6:** Powder X-ray diffraction of $Nd_{0.9}Sr_{0.1}NiO_2$ (a) and $Nd_{0.8}Sr_{0.2}NiO_2$ (b) and the corresponding Pawley refinement. The black dots are the observed data, the red line is the calculated fit, and gray line shows the difference between the two. Positions of allowed reflections are indicated by vertical red ($Nd_{1-x}Sr_xNiO_2$) and black (Ni) lines.

**Electrical transport properties**

The temperature dependence of the resistivity of $Nd_{1-x}Sr_xNiO_3$ ($x = 0.1, 0.2, 0.3$) and $Nd_{0.8}Sr_{0.2}NiO_2$ is shown in Fig. 7. It can be seen that $NdNiO_3$ shows the characteristic first-order phase transition from a high-temperature paramagnetic metal to a low temperature antiferromagnetic insulator and that this transition is suppressed by Sr doping. These results are in good agreement with the results obtained in Ref. [8]. We display the result of $Nd_{0.8}Sr_{0.2}NiO_2$ only as a representative example of the temperature-dependence of the measured resistivity. Although it matches well with the report of Q. Li *et al.* [20], we do not believe that this is truly representative of the intrinsic resistance of $Nd_{0.8}Sr_{0.2}NiO_2$. The issue here is that the temperature dependence is uncharacteristic of an oxide – typically if the actual room temperature resistivity is on the order of an Ω-cm at room temperature, as we measure here, then the oxide will display semiconducting resistivity with a much higher low temperature resistance than measured. The reduction process yields fragile pellets with low density, and unfortunately post-reduction annealing at higher temperatures causes the infinite layer phase to decompose. Regardless, we are unable to see any hint of superconductivity in our infinite layer samples, but we think that the intrinsic resistivity remains an open question.

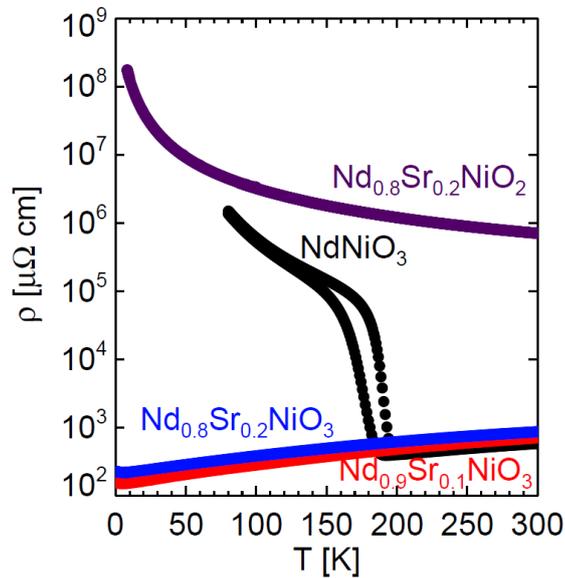

**Fig. 7:** Temperature dependent resistivity of the $Nd_{1-x}Sr_xNiO_3$ ($x$ = 0.0, 0.1 and 0.2) and $Nd_{1-x}Sr_xNiO_2$ ($x$ = 0.2) samples.

**Neutron diffraction**

As described above, powder neutron diffraction measurements were performed on a sample of $Nd_{0.9}Sr_{0.1}NiO_2$ that had been synthesized from acetates. Measurements were taken at both 3 K and 300 K and are shown in Fig. 8(a) and (b), respectively. Close inspection of the powder patterns revealed nonpeak splitting or the appearance of peaks at 3 K that were not visible at 300 K; thus, we observed neither a structural phase transition nor evidence of long-range magnetic order. In fact, we were able to obtain reasonable fits to the data at both temperatures using a combination of the *P4/mmm* space group for the 112 phase with an additional fcc nickel second phase. The calculated patterns as well as the difference plots are shown for each of the refinements in 8(a) and 8(b), and the refined parameters are listed in Table 2.

It is useful to compare the results here with those of Hayward and Rosseinsky on $NdNiO_{2+x}$ [17] with the caveat that their sample contained no Sr. Similar to the present work, they refined their neutron powder data in space group *P4/mmm* both at high (290 K) and low temperature (1.7 K). In the case of Hayward's structural refinement, a molar fraction of 18% Ni was refined, which is significantly higher than that found in the present case (5.4 %). It is possible that this may result from the differing reduction procedure applied (different reduction agents, different compositions, different temperature profiles), but it may also be highly sample-dependent. To improve their fit to the data, Hayward *et al.* introduced a second phase with apical oxygens located in the Nd layer and significant relaxation of the atomic positions vis-à-vis the stoichiometric phase. We found that satisfactory refinement of our $Nd_{0.9}Sr_{0.1}NiO_2$ data did not necessitate addition of this phase. This may very well relate to the composition, as we expect there to be an increased propensity for interstitial apical oxygens to appear as the nickel valence is reduced towards 1+.

In addition, Hayward *et al.* observed no evidence of antiferromagnetic ordering. Assuming a *G*-type antiferromagnetic structure, they estimated an upper limit of 0.06 $\mu_B$/Ni as their sensitivity limit. The sensitivity of our measurement at 3 K to a hypothetical *G*-type antiferromagnetic structure is significantly poorer, on the order of 0.5 $\mu_B$. Thus, neither the parent compound nor the 10% Sr doped compound shows evidence of magnetic order within these estimated sensitivity limits. When considering the lack of observed antiferromagnetic order within the context of the superconductivity observed in thin films, we note that the strained, epitaxial films may exhibit different intrinsic magnetic behavior.

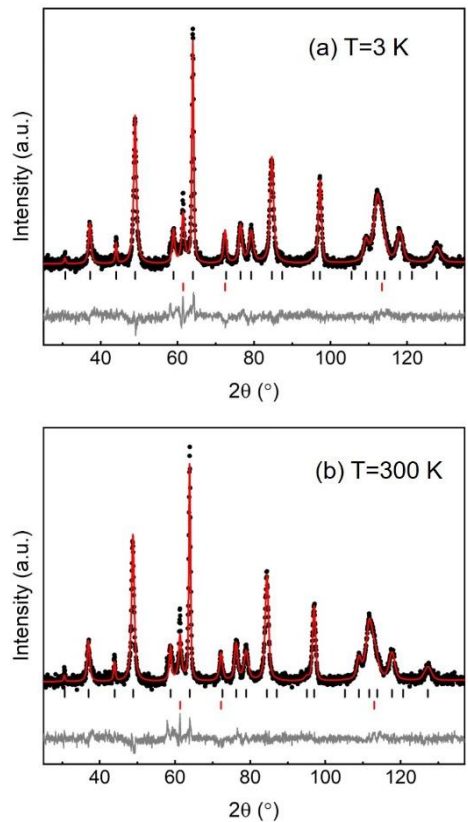

**Fig. 8:** Neutron diffraction and the corresponding Rietveld refinement of $Nd_{0.9}Sr_{0.1}NiO_2$ at 3 K (a) and at 300 K (b). The black dots are the observed data, the red line is the calculated fit, and gray line shows the difference between the two. Positions of allowed reflections are indicated by vertical black (Nd$_{1-x}$Sr$_x$NiO$_2$) and red (Ni) lines.

**Table 2** Refined crystal structural parameters of Nd$_{0.9}$Sr$_{0.1}$NiO$_2$ at 3 K and 300 K. *a* and *c* are the lattice parameters of the tetragonal crystal, *V* is volume of unit cell; $U_{iso}$ is the isotropic atomic displacement parameter. The quality of the agreement between the observed and calculated profiles is measured by various R factors.

|  |  | 3 K | 300 K |
|---|---|---|---|
| *a (Å)* |  | 3.9132 (1) | 3.9206 (1) |
| *c (Å)* |  | 3.2539(3) | 3.2664 (3) |
| *V (Å³)* |  | 49.828(5) | 50.208 (5) |
| *c/a* |  |  |  |
| Nd | coordinate | (0.5,0.5,0.5) | (0.5,0.5,0.5) |
|  | $100 \times U_{iso}$ *(Å²)* | 1.25(6) | 1.19(7) |
| Sr | coordinate | (0.5,0.5,0.5) | (0.5,0.5,0.5) |
|  | $100 \times U_{iso}$ *(Å²)* | 1.25(6) | 1.19(7) |
| Ni | coordinate | (0,0,0) | (0,0,0) |
|  | $100 \times U_{iso}$ *(Å²)* | 2.06(5) | 1.77(6) |
| O | coordinate | (0.5,0,0) | (0.5,0,0) |
|  | $100 \times U_{iso}$ *(Å²)* | 1.56(5) | 1.45(6) |
| *Nd-O length (Å)* |  | 2.5444 (1) × 8 | 2.5513 (2) × 8 |
| *Ni-O length (Å)* |  | 1.9565(6) × 4 | 1.9602 (1) × 4 |
| *wRp/Rp* (%) |  | 4.41/3.46 | 4.24/3.34 |

## Conclusion

In this work, high quality bulk polycrystalline Nd$_{1-x}$Sr$_x$NiO$_3$ (*x* = 0, 0.1 and 0.2) samples have been synthesized using a combined sol-gel combustion and high pressure annealing technique. The evolution from NdNiO$_3$ to NdNiO$_2$ was monitored using *in-situ* synchrotron X-ray diffraction during the reduction process. Nd$_{1-x}$Sr$_x$NiO$_3$ (*x* = 0.1 and 0.2) samples were reduced to Nd$_{1-x}$Sr$_x$NiO$_2$ (*x* = 0.1 and 0.2) using calcium hydride in a sealed evacuated tube. Our bulk infinite-layer samples showed no sign of superconductivity, though measurements of the intrinsic resistivity are hampered by loose polycrystalline compaction as well as metallic nickel impurities. Neutron diffraction data of the hole-doped infinite layer phase could be well fit by a combination of the *P4/mmm* space group with a nickel impurity. Within the limits of our powder diffraction experiment, no evidence of magnetic ordering down to 3 K was observed.

## Acknowledgements

Work in the Materials Science Division at Argonne National Laboratory was supported by the U.S. Department of Energy, Office of Science, Basic Energy Sciences, Materials Science and Engineering Division. The identification of any commercial product or trade name does not imply endorsement or recommendation by the National Institute of Standards and Technology.